# Security Risks Concerns of Generative AI in the IoT

Honghui Xu, Yingshu Li, Olusesi Balogun, Shaoen Wu, Yue Wang, Zhipeng Cai

*Abstract*—In an era where the Internet of Things (IoT) intersects increasingly with generative Artificial Intelligence (AI), this article scrutinizes the emergent security risks inherent in this integration. We explore how generative AI drives innovation in IoT and we analyze the potential for data breaches when using generative AI and the misuse of generative AI technologies in IoT ecosystems. These risks not only threaten the privacy and efficiency of IoT systems but also pose broader implications for trust and safety in AI-driven environments. The discussion in this article extends to strategic approaches for mitigating these risks, including the development of robust security protocols, the multi-layered security approaches, and the adoption of AI technological solutions. Through a comprehensive analysis, this article aims to shed light on the critical balance between embracing AI advancements and ensuring stringent security in IoT, providing insights into the future direction of these intertwined technologies.

## I. INTRODUCTION

THE intersection of generative Artificial Intelligence (AI) and the Internet of Things (IoT) marks a transformative era in technology, redefining the boundaries of innovation and application. Generative AI, which is a subset of AI focused on creating synthetic instances of data to mimic real-world patterns, is revolutionizing the operation way in IoT systems. From enhancing user experiences in smart homes [1], [2] to predictive maintenance in industrial settings [3], the capabilities of generative AI in IoT are expansive and profoundly impactful. However, the integration of these advanced technologies raises significant security concerns that cannot be overlooked [4], [5], [6].

IoT devices are becoming ubiquitous in various fields, including critical ones like healthcare, finance, and transportation. Since IoT devices often operates on the fringe of traditional security measures and generates vast amounts of data, these devices are easy to be attacked by the cybercriminals. For example, the botnet attack turned a large number of Internet-connected devices into a botnet. And many other high-profile incidents also underline the vulnerability of IoT ecosystems to security breaches, which not only disrupt services but also compromise the privacy and safety of individuals and organizations [7], [8]. Hence, exploring these security risks of generative AI in IoT is urgent and of necessity in the increasing sophistication and prevalence of cyber threats in an ever-more connected world.

In this article, we aim to provide an in-depth analysis of the security risks associated with the use of generative AI in IoT. Although generative AI brings plentiful benefits in enhancing the functionality and efficiency of IoT systems, there is a pressing need to understand and address the potential security pitfalls that accompany this technological synergy between generative AI and IoT. This article expects to offer a comprehensive perspective that not only outlines the challenges but also delves into potential strategies for mitigating these risks. This analysis is crucial for stakeholders in the IoT ecosystem, including technology developers, policymakers, and end-users, to ensure that the deployment of generative AI in IoT is secure and reliable.

The rest of this article is organized as follows. In Section II, we present the overview of generative AI in IoT. After detailing the analysis of security risks in Section III, we discuss the strategies for mitigating risks in Section IV. In Section V, we provide insights into the future direction of generative AI in IoT. Finally, we end up with a conclusion in Section VI.

## II. OVERVIEW OF GENERATIVE AI IN IOT

Generative AI encompasses sophisticated algorithms designed to produce synthetic data that closely resembles real-world patterns. In the realm of IoT systems, a network of interconnected devices has the capability to communicate and process data autonomously. In today's technological landscape, generative AI is driving IoT towards unprecedented levels of innovation and efficiency, opening up new possibilities for smart and automated solutions. In the following, we will provide a comprehensive overview of generative AI within the IoT framework, focusing on four aspects, including applications, advantages, weaknesses, and security concerns.

### A. Applications

There are various generative AI's applications in IoT. For example, in smart cities, the generative AI enables the simulation of complex urban environments to optimize traffic flow and energy consumption [1], [2]; in healthcare, the generative AI helps in generating synthetic patient data for research while preserving privacy [9], [10]; and in industrial IoT, generative AI-driven predictive maintenance effectively reduces downtime by anticipating and identifying equipment malfunctions before they occur [3]. These applications underscore the generative AI's potential to enhance efficiency, accuracy, and innovation in various sectors.

### B. Advantages

The advantages of employing generative AI within IoT systems are manifold. Firstly, the generation AI allows for the generation of vast and varied datasets for training machine learning models where real-world data might be scarce or sensitive. This synthetic data generation is crucial in fields where data privacy is paramount. Secondly, generative AI

Honghui Xu and Shaoen Wu are with Department of Information Technology, Kennesaw State University, USA
Yingshu Li, Olusesi Balogun, Yue Wang, and Zhipeng Cai are with Department of Computer Science, Georgia State University, USA

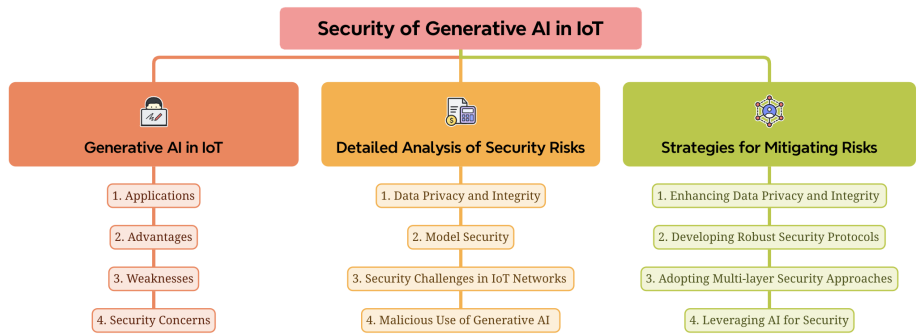

Fig. 1. Security Risks Concerns of Generative AI in the IoT

enables better decision-making and forecasting. By simulating different scenarios by using generative AI, IoT systems can predict outcomes and optimize processes, leading to increased operational efficiency. Lastly, generative AI drives innovation by pushing the boundaries of what IoT systems can achieve, paving the way to create new products and services.

*C. Weaknesses*

The integration of generative AI into IoT has some weaknesses as well. The sophistication of generative AI models means they require substantial computational resources, which will be a challenge, particularly in IoT environments where devices may have limited processing capabilities. Besides, the accuracy of the data generated by these AI models is crucial since inaccuracies can lead to flawed decision-making, especially in critical applications like healthcare. Additionally, integrating generative AI into existing IoT ecosystems may pose data heterogeneity challenges. Specifically, when dealing with heterogeneous devices and data formats, ensuring seamless integration and compatibility across diverse IoT components can be complex and may require additional resources and expertise.

*D. Security Concerns*

The most pressing concern, however, lies in the realm of security. Generative AI algorithms are capable of producing highly realistic data, which can be a double-edged sword. On one hand, this capability is invaluable for creating diverse datasets for training and simulation. On the other hand, it can be exploited to generate sophisticated cyber-attacks, such as deepfakes or realistic phishing content, posing significant threats to the integrity and security of IoT systems. Moreover, the AI models themselves can become targets for cyber-attacks. For example, adversarial attacks can lead to erroneous outputs from AI models, causing systemic failures in IoT operations. Furthermore, the interconnected nature of IoT devices amplifies these risks, where means that a security breach in one device can potentially compromise an entire network. That is to say, with an ever-growing number of devices in IoT systems, it will be more challenging to establish robust security protocols that can keep pace with emerging cyber threats.

III. DETAILED ANALYSIS OF SECURITY RISKS

The advent of generative AI in the realm of the IoT represents a groundbreaking fusion of two cutting-edge technologies. However, beneath the surface of these technological advancements lies a complex web of security risks. These risks are not only multifaceted but also magnified by the intrinsic properties of Generative AI and the extensive, interconnected nature of IoT networks. In the following, we will elaborate a detailed analysis of these security risks, it is crucial to understand that the implications extend beyond mere technical challenges. This analysis aims to unravel these risks, providing insights into their nature, potential impact, and the urgency with which they need to be addressed.

*A. Data Privacy and Integrity*

Data privacy and integrity emerge as paramount concerns in the dynamic intersection of generative AI and IoT. The crux of the issue lies in the capability of generative AI to create and manipulate vast amounts of data. However, if not properly safeguarded, this technology can lead to significant breaches in privacy and compromises in data integrity [11], [12]. Especially, in IoT environments, where devices continuously collect and transmit data, the potential for such data leakage is exponentially magnified. Moreover, the interconnected nature of IoT devices indicates that a single point of vulnerability can have cascading effects, spreading compromised data across the network. This not only jeopardizes the privacy of individual users but also undermines the trustworthiness of entire IoT systems. For instance, a breach in a single smart home device could potentially expose a plethora of personal data, from daily routines to sensitive personal information. As these systems grow increasingly sophisticated, the techniques used to exploit their vulnerabilities will also become more formidable. Therefore, we need a proactive and constantly evolving approach to securing data, ensuring both its privacy and integrity.

*B. Model Security*

The security of generative AI models is also a critical concern in IoT ecosystems. These models, the core of AI-driven systems, are susceptible to various forms of compromise, which can severely undermine the functionality and safety of IoT applications [13]. One prevalent form of attack is the theft of AI models. Cybercriminals may steal these models to understand their structure and functioning, enabling them to craft more effective attacks or develop countermeasures against AI-driven security systems. Another significant threat is the poisoning of AI models during their training phase. In this scenario, attackers inject malicious data into the training set, causing the model to learn incorrect patterns or biases. This could result in compromised decision-making when the model is implemented in real-world IoT applications. For instance, a compromised AI model in a smart security system might fail to recognize genuine security threats, leaving the system vulnerable. These risks underscore the importance of securing AI models against unauthorized access and manipulation.

*C. Security Challenges in IoT Networks*

The inherent complexity and diversity of the IoT networks amplify the security challenges posed by the integration of Generative AI. Each device within an IoT ecosystem, from simple sensors to complex processors, represents a potential vulnerability point. A breach in a single device can have a domino effect, potentially compromising the entire network. Besides, the heterogeneity of IoT devices, each of which varies capabilities and security protocols, poses a unique challenge. Standardizing security measures across such a diverse range of devices is difficult, often leading to inconsistencies in security implementations. These inconsistencies can create security gaps, leaving the network vulnerable to attacks. In a word, the key challenge in securing IoT networks lies in their extensive, interconnected nature. This interconnectedness demands a robust and comprehensive security strategy that extends beyond individual devices to encompass the entire network.

*D. Malicious Use of Generative AI*

When these abilities are used for malicious purposes within the IoT ecosystem, the advanced capabilities of generative AI in creating highly realistic and convincing data present significant security concerns as well [14]. The creation of sophisticated deepfakes or the fabrication of misleading data using generative AI technologies poses a grave threat to the integrity and reliability of IoT systems. These AI-generated forgeries, which can be audio, video, or other data types, are becoming increasingly indistinguishable from authentic content. In the context of IoT, where devices often rely on data inputs to make automated decisions, deepfakes could be used to mislead systems. For example, audio deepfakes could be used to impersonate voice commands to smart home devices, leading to unauthorized access or triggering of actions that compromise user safety and privacy. Moreover, generative AI can be exploited to fabricate data streams or sensor outputs in IoT environments, leading to manipulated behaviors in connected systems. This can have far-reaching consequences, especially in critical applications such as healthcare monitoring systems or industrial automation, where decision-making relies heavily on the accuracy and authenticity of data. As such, the potential for misuse of generative AI in IoT necessitates robust countermeasures, including advanced detection techniques and stringent security protocols, to protect against these evolving and sophisticated threats.

To sum up, while generative AI offers tremendous potential in enhancing IoT applications, it is also imperative to acknowledge the associated security risks. These risks, ranging from data privacy concerns to vulnerabilities in AI algorithms, pose significant challenges to the safe and effective deployment of these technologies. As the IoT continues to evolve, a proactive and comprehensive approach to security is essential to ensure the trustworthiness and reliability of these interconnected systems. The future development of generative AI and IoT must prioritize security to fully realize their potential.

## IV. STRATEGIES FOR MITIGATING RISKS

As mentioned above, although generative AI is propelling the IoT towards new heights of innovation and efficiency, it also brings with it a set of challenges, particularly in terms of security. Addressing these challenges is critical to harness the full potential of Generative AI in IoT and ensure the safe, reliable operation of these interconnected systems. Therefore, the research focus must be on developing strategies that not only leverage the capabilities of generative AI but also safeguard against its potential vulnerabilities. In the following, we explore various strategies for reducing the security risks inherent in the integration of generative AI within IoT ecosystems.

*A. Enhancing Data Privacy and Integrity*

In the rapidly evolving landscape of generative AI within the IoT, the protection of critical data is paramount. Because data is the cornerstone of AI functionality since it drives the learning processes and decision-making capabilities of these advanced systems. To enhance data privacy and integrity, a multifaceted approach is essential, incorporating techniques like encryption, anonymization, stringent access control, and continuous monitoring. (1) Encryption serves as a first line of defense, which can encode the data in such a way that it can only be accessed and interpreted by authorized entities. Advanced encryption methods are particularly vital in IoT environments, where data is frequently transmitted across networks and devices. By ensuring that the data remains encrypted during transit, we can significantly reduce the risk of unauthorized access or interception. (2) Anonymization also plays an important role in protecting user privacy. Techniques like differential privacy enable the extraction of useful information from datasets while obscuring individual identities. These techniques are crucial in maintaining the delicate balance between data utility and personal privacy. (3) Stringent access control is another critical component of a robust data protection strategy. These controls guarantee that only authorized personnel have access to sensitive data, thus minimizing the risk of accidental or malicious data breaches. This can include multi-factor authentication, role-based access controls, and regular audits of access logs to detect any unauthorized or suspicious activities. (4) Continuous monitoring and updating of these security measures are vital to adapt to new threats and vulnerabilities in the evolving digital landscape. By fostering a sense of security awareness, organizations can further strengthen their defenses against potential threats to their AI-driven IoT systems.

### B. Developing Robust Security Protocols

The evolving nature of cyber threats necessitates regular updates and patches as a fundamental aspect of securing IoT devices and AI models, which means that cybersecurity is no longer static, thus requiring dynamic and proactive measures. Automated update mechanisms play a critical role in this process, continuously ensuring that devices and systems operate on the most current software versions, which include the latest security enhancements to protect against emerging vulnerabilities. Moreover, robust security protocols that can encompass the entire lifecycle of IoT devices and generative AI models are also able to mitigate the effective risk. These protocols are integral to the development process, including secure coding practices and regular security audits, which can guarantee that security considerations are embedded from the onset, thereby fortifying the devices and systems against potential threats. In a nutshell, these comprehensive security protocols are extremely important for maintaining the integrity and reliability of IoT and generative AI technologies in the face of constantly changing cyber threats.

### C. Adopting Multi-Layered Security Approaches

A multi-layered, or defense-in-depth, approach to security is beneficial and essential in the realm of IoT ecosystems [15]. This multi-layered security system is designed such that if one layer fails or is compromised, subsequent layers continue to provide protection. This strategy involves implementing multiple tiers of defense, creating a security system that is far more resilient than any single-layered approach. At the most basic level, this could include deploying firewalls, which act as the first line of defense against external attacks, controlling the incoming and outgoing network traffic based on predetermined security rules. In addition to firewalls, intrusion detection systems (IDS) can be leveraged to monitor network traffic for suspicious activity and potential threats, alerting the network administrators to any anomalies. Besides, regular vulnerability assessments are also another key component of a multi-layered security approach, which is helpful in identifying and addressing potential security gaps in the system before they can be exploited by attackers. Additionally, the multi-layered security approach extends beyond just technical measures, which can encompass administrative controls such as security policies, user training, and incident response plans. What's more, physical security measures, such as secure access to hardware, also form an integral part of a multi-layered approach. In a word, the multi-layered approach involves a comprehensive and strategic blend of technical, administrative, and physical security measures, all working to protect the vast and varied components of the IoT landscape from ever-evolving cyber threats.

### D. Leveraging AI for Security

Although the rise of AI introduces new complexities into the IoT landscape, AI itself can be applied as a powerful tool in the battle against security risks. Machine learning algorithms, a cornerstone of AI, can analyze vast datasets far more efficiently than traditional methods. (1) These algorithms are good at detecting unusual patterns or anomalies in network traffic, which are often indicators of security threats. By continuously learning from the data they process, these systems become increasingly adept at identifying potential risks, sometimes even as they emerge. This real-time detection is crucial for modern networks, where threats can evolve rapidly and traditional security measures may lag behind. (2) Beyond detection, AI can actively engage in threat mitigation. For instance, upon identifying a potential security breach, AI systems can initiate protocols to isolate the affected network segment, thereby limiting the spread of the attack. This immediate response can be invaluable in minimizing damage while human experts assess and address the situation. (3) AI's predictive capabilities can also simulate various cyber-attack scenarios to help organizations to rigorously test their defenses. These simulations can reveal vulnerabilities that might not be apparent during standard security audits, allowing for preemptive strengthening of defenses, especially in an era where cyber threats are not only frequent but also increasingly sophisticated. (4) AI's role in cybersecurity can extend to enhancing other security measures as well. For instance, in the realm of identity verification, AI can analyze behavioral patterns to detect anomalies, adding an extra layer of authentication. In encryption, AI can manage complex encryption keys more efficiently, ensuring data remains secure yet accessible to authorized users. In essence, by leveraging AI's ability to process and analyze data at unprecedented scales, we can transform the reactive nature of traditional security into a dynamic and adaptive defense system through the integration of AI into traditional cybersecurity strategies.

In summary, mitigating the security risks associated with the use of generative AI in IoT requires a comprehensive approach. We can help organizations improve the security posture of their IoT ecosystems by enhancing data privacy and integrity, developing robust security protocols, adopting multi-layered security approaches, and leveraging AI itself for security. However, the limitations of the proposed strategies for mitigating risks in the integration of generative AI and IoT still present considerable challenges. (1) Enhancing data privacy and integrity will introduce burdensome overheads for smaller organizations. In particular, anonymization and encryption are not infallible, and maintaining privacy in dynamic data environments is an ongoing struggle. (2) Managing robust security protocols can still leave systems vulnerable when facing the rapid evolution of threats and the inherent complexities, and human error also remains a significant weak point in the implementation of security protocols. (3) The adoption of multi-layered security approaches brings its own set of complications. To be specific, managing these layers can be intricate and resource-intensive, and can lead to conflicts or redundancies, potentially offering a false sense of security.

(4) Leveraging AI for security is heavily dependent on the quality of data and requires significant expertise and resources. These limitations underscore the need for a dynamic, adaptive approach to security. The ongoing investment in security are essential to safeguarding the potential of generative AI and IoT integrations.

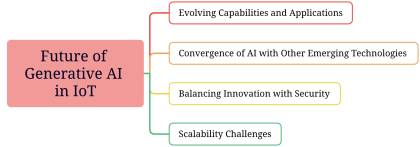

Fig. 2. Future of Generative AI in IoT

## V. Future of Generative AI in IoT

As we look towards the future, the fusion of generative AI with the IoT still presents a landscape brimming with both potential and challenges. This discussion deeply investigates the anticipated future developments of generative AI in IoT.

**Evolving Capabilities and Applications:** The continuous advancement in generative AI algorithms promises to expand the capabilities and applications within IoT. We can anticipate more sophisticated data generation, improved predictive analytics, and enhanced automation in various sectors. For instance, in smart cities, generative AI could lead to more efficient urban planning and management systems. In healthcare, personalized patient care through AI-driven diagnostics and treatment plans could become a norm. In industry, the generative AI can help enhanced automation, predictive maintenance, and optimization of manufacturing processes in order to realize improved efficiency and reduced downtime. However, these advancements will invariably introduce new complexities and challenges, particularly in terms of security considerations.

**Convergence of AI with Other Emerging Technologies:** The future will likely see generative AI not just in isolation but in convergence with other emerging technologies such as quantum computing, 5G, edge computing, and reinforcement learning. This convergence has the potential to exponentially increase the capabilities of IoT systems but also complicates the security landscape. Generative AI will play a crucial role in the next industrial revolution, often termed Industry 4.0. Industry 4.0's integration with IoT in industrial settings will lead to smarter, more efficient manufacturing processes. This integration also means that the stakes for security breaches will be higher, potentially impacting critical infrastructure and economies. What's more, the adaptive nature of reinforcement learning algorithms used in the generative AI may lead to unforeseen vulnerabilities and adversarial attacks.

**Balancing Innovation with Security:** As generative AI becomes more powerful, the security risks it poses will likely evolve and become more sophisticated. The future will demand a delicate balance between embracing the innovative capabilities of generative AI and ensuring robust security measures are in place. This balance will require continuous research and development in security technologies, as well as dynamic regulatory frameworks that can adapt to the evolution of these generative AI and IoT technologies. The future will require proactive strategies to solve new types of cyber-attacks and vulnerabilities.

**Security Risk of Multi-Agent Generative AI in IoT:** The proliferation of multi-agent generative AI within IoT introduces security concerns, as the decentralized structure of multi-agent systems can cause vulnerabilities in agent communication and coordination. These susceptibilities present opportunities for malicious attackers to exploit, potentially leading to unauthorized access, data manipulation, or system disruption. Furthermore, the dynamic and adaptive characteristics inherent in multi-agent systems heighten the challenge of promptly detecting and addressing security threats, posing substantial risks to the integrity, confidentiality, and availability of IoT data and services.

**Scalability Challenges:** As IoT networks continue to expand, managing and securing these vast networks will become increasingly challenging. In terms of the application of generative AI in the IoT, the scalability challenges will be caused by the increasing volume and complexity of data generated by IoT devices. In the meantime, for the future developments, we should also focus on scalable security solutions that can adapt to the growing size and complexity of IoT networks.

In summary, the future of generative AI in IoT will experience continuous innovation and evolution. To be specific, by adopting a proactive and collaborative approach, we can harness the full potential of Generative AI in IoT, creating a future that is technologically advanced. Nevertheless, when preparing for the integration of these technologies into many real applications, we will also require an effort to address the emerging security challenges.

## VI. Conclusion

As we have explored throughout this article, the integration of generative AI into IoT presents profound implications for the future of technology and society. The utilization of generative AI transforms IoT applications into various sectors, offering unprecedented levels of efficiency, automation, and data-driven insights. However, this fusion also brings to the forefront significant security challenges that must be addressed with urgency. The security risks associated with generative AI in IoT are multifaceted, ranging from data privacy breaches to the potential misuse of AI technologies. These risks not only pose threats to the integrity and reliability of IoT systems but also to the trust and safety of users. Therefore, it is imperative that as we advance in developing these technologies, and we should equally emphasize on developing robust security measures to safeguard against potential vulnerabilities. In the end of this article, we delve into the future of generative AI in IoT with cautious optimism by proposing approaches that encompasses technological innovation and collaborative efforts. In the meantime, we think that a balanced approach is necessary, where the excitement of technological advancement is tempered with a steadfast commitment to security considerations. In conclusion, the convergence of generative AI and IoT marks a new frontier in technology. As we embrace this future, we need to keep an eye on security and a firm commitment to ensuring a safe and reliable technological landscape.

ACKNOWLEDGEMENT

This work was partly supported by the National Science Foundation of U.S. (2413622, 2244219, 2231209, 2315596, 2146497).

## REFERENCES

[1] M. E. E. Alahi, A. Sukkuea, F. W. Tina, A. Nag, W. Kurdthongmee, K. Suwannarat, and S. C. Mukhopadhyay, "Integration of iot-enabled technologies and artificial intelligence (ai) for smart city scenario: Recent advancements and future trends," *Sensors*, vol. 23, no. 11, p. 5206, 2023.

[2] B. Yamini, M. Jayaprakash, S. Logesswari, V. Ulagamuthalvi, R. Porselvi, and G. Uthayakumar, "Enhanced expectation-maximization algorithm for smart traffic iot systems using deep generative adversarial networks to reduce waiting time," in *2023 4th International Conference on Electronics and Sustainable Communication Systems (ICESC)*. IEEE, 2023, pp. 380–385.

[3] X. Liang, Z. Liu, H. Chang, and L. Zhang, "Wireless channel data augmentation for artificial intelligence of things in industrial environment using generative adversarial networks," in *2020 IEEE 18th International Conference on Industrial Informatics (INDIN)*, vol. 1. IEEE, 2020, pp. 502–507.

[4] Z. Xiong, W. Li, and Z. Cai, "Federated generative model on multi-source heterogeneous data in iot," in *Proceedings of the AAAI Conference on Artificial Intelligence*, vol. 37, no. 9. AAAI, 2023, pp. 10 537–10 545.

[5] S. De, M. Bermudez-Edo, H. Xu, and Z. Cai, "Deep generative models in the industrial internet of things: a survey," *IEEE Transactions on Industrial Informatics*, vol. 18, no. 9, pp. 5728–5737, 2022.

[6] S. Malik, A. K. Tyagi, and S. Mahajan, "Architecture, generative model, and deep reinforcement learning for iot applications: Deep learning perspective," *Artificial Intelligence-based Internet of Things Systems*, pp. 243–265, 2022.

[7] N. Abdalgawad, A. Sajun, Y. Kaddoura, I. A. Zualkernan, and F. Aloul, "Generative deep learning to detect cyberattacks for the iot-23 dataset," *IEEE Access*, vol. 10, pp. 6430–6441, 2021.

[8] M. A. Ferrag, M. Debbah, and M. Al-Hawawreh, "Generative ai for cyber threat-hunting in 6g-enabled iot networks," *arXiv preprint arXiv:2303.11751*, 2023.

[9] K. Nova, "Generative ai in healthcare: advancements in electronic health records, facilitating medical languages, and personalized patient care," *Journal of Advanced Analytics in Healthcare Management*, vol. 7, no. 1, pp. 115–131, 2023.

[10] C. Dilibal, B. L. Davis, and C. Chakraborty, "Generative design methodology for internet of medical things (iomt)-based wearable biomedical devices," in *2021 3rd International Congress on Human-Computer Interaction, Optimization and Robotic Applications (HORA)*. IEEE, 2021, pp. 1–4.

[11] M. Gupta, C. Akiri, K. Aryal, E. Parker, and L. Praharaj, "From chatgpt to threatgpt: Impact of generative ai in cybersecurity and privacy," *IEEE Access*, 2023.

[12] T. Wang, Y. Zhang, S. Qi, R. Zhao, Z. Xia, and J. Weng, "Security and privacy on generative data in aigc: A survey," *arXiv preprint arXiv:2309.09435*, 2023.

[13] S. A. Khowaja, K. Dev, N. M. F. Qureshi, P. Khuwaja, and L. Foschini, "Toward industrial private ai: A two-tier framework for data and model security," *IEEE Wireless Communications*, vol. 29, no. 2, pp. 76–83, 2022.

[14] Z. Xiong, H. Xu, W. Li, and Z. Cai, "Multi-source adversarial sample attack on autonomous vehicles," *IEEE Transactions on Vehicular Technology*, vol. 70, no. 3, pp. 2822–2835, 2021.

[15] R. E. Crossler, F. Bélanger, and D. Ormond, "The quest for complete security: An empirical analysis of users' multi-layered protection from security threats," *Information Systems Frontiers*, vol. 21, pp. 343–357, 2019.

**Honghui Xu** (hxu16@gsu.edu) received a Ph.D. degree in computer science from Georgia State University, Atlanta, GA, USA, in 2023 and a Bachelor's degree from University of Electronic Science and Technology of China. He is currently an Assistant Professor of Information Technology with Kennesaw State University, Kennesaw, GA, USA. His research focuses on data privacy and security in AI, deep learning, and Internet of Things. He served as the Program Committees for AAAI, SDM, and ICMC, and was a recipient of the Best Paper Award of the IEEE SmartData 2022.

**Yingshu Li** (yili@gsu.edu) received her Ph.D. and M.S. degrees from the Department of Computer Science and Engineering at University of Minnesota-Twin Cities. Dr. Li is currently a Professor in the Department of Computer Science and an affiliated faculty member in the INSPIRE Center at Georgia State University. Her research interests include Privacy-aware Computing, Management of Big Sensory Data, Internet of Things, Social Networks, and Wireless Networking. Dr. Li is the recipient of the NSF CAREER Award. Dr. Li was nominated to the 2022 N²Women: Stars in Computer Networking and Communications.

**Olusesi Balogun** (obalogun6@student.gsu.edu) received the B.S. degree in computer engineer- ing from Obafemi Awolowo University, Nigeria, in 2014. He is currently pursuing the Ph.D. degree with the Department of Computer Science, Georgia State University, Atlanta, Georgia, USA. His research interests include insider threat detection, moving target defense, access control models, and security and privacy in cyber-physical systems (CPS). He is a Student Member of ACM and IEEE.

**Shaoen Wu** (swu10@kennesaw.edu) is a Senior Member of IEEE and received the Ph.D. degree in computer science from Auburn University, Auburn, AL, USA, in 2008. He is currently a Chair Professor of Information Technology with Kennesaw State University, Kennesaw, GA, USA. His current research interests include Internet of Things, wireless and mobile networking, cybersecurity, and cyber–physical systems. Prof. Wu was a recipient of the Best Paper Awards of the IEEE Globecom 2019, IEEE ISCC 2008, and the ANSS 2011. He has served on the chairs and the committees for various conferences, such as the IEEE INFOCOM, ICC, and Globecom, and an editor for several journals, including IEEE TRANSACTION ON MULTIMEDIA. He currently serves as the Vice Chair of North America for the IEEE ComSoc Multimedia Communication Committee (MMTC).

**Yue Wang** (ywang182@gsu.edu) is a senior member of IEEE and received the Ph.D. degree in communication and information system from the Beijing University of Posts and Telecommunications (BUPT), Beijing, China, in 2011. He is currently an Assistant Professor with the Department of Computer Science, Georgia State University, Atlanta, GA, USA. Previously, he was a Research Assistant Professor with the Department of Electrical and Computer Engineering, George Mason University, Fairfax, VA, USA. His general interests lie in the interdisciplinary areas of machine learning, signal processing, wireless communications, and their applications in cyber-physical systems. His specific research focuses on trustworthy AI, compressive sensing, massive MIMO, millimeter-wave communications, wideband spectrum sensing, cognitive radios, direction of arrival estimation, high-dimensional data analysis, and distributed optimization and learning.

**Zhipeng Cai** (zcai@gsu.edu) is a Fellow of IEEE and received his PhD and M.S. degrees in the Department of Computing Science at University of Alberta. He is currently a Professor in the Department of Computer Science at Georgia State University, and also an Affiliate Professor in the Department of Computer Information System at Robinson College of Business, as well as a Director at INSPIRE Center. His research expertise lies in the areas of Resource Management and Scheduling, High Performance Computing, Cyber-Security, Privacy, Networking, and Big Data. He is an Editor-in-Chief for WCMC, and an Associate Editor-in-Chief for Elsevier HCC Journal, as well as an editor for various journals, such as IEEE TKDE, IEEE TVT, IEEE TWC, and IEEE TCSS. He also serves as a Steering Committee Co-Chair for WASA, and a Steering Committee member for COCOON, IPCCC and ISBRA. He is the recipient of an NSF CAREER Award.